\documentclass[aps,prl,twocolumn,superscriptaddress]{revtex4-1}
\usepackage{graphicx}
\usepackage{caption}
\DeclareCaptionLabelSeparator{dot}{. }
\makeatletter
\def\justified{
	\let\\\@normalcr
	\@rightskip\z@skip \rightskip\@rightskip
	\leftskip\z@skip
	\parindent 0em\relax
	\setlength{\parfillskip}{0pt plus 1fil}}
\DeclareCaptionJustification{justified}{\justified}
\usepackage{subcaption}
\usepackage{amsmath}
\usepackage{braket}
\usepackage{units}
\usepackage{ragged2e}
\usepackage[colorlinks,urlcolor=blue ,citecolor=blue ,linkcolor=blue ]{hyperref}
\usepackage{xcolor}
\usepackage{nicefrac} % for \nicefrac macro
\usepackage{lipsum}
\usepackage{nameref}
\usepackage{hyperref}
\usepackage{textcomp} %for \textmu
\usepackage{urwchancal}
\usepackage{upgreek} %for \upmu
\usepackage{amsmath} % for \multiline

\newcommand{\Er}{\ensuremath{^{166}}{\rm Er} }

\newcommand{\as}{\ensuremath{a_{\rm s}}}

\newcommand{\kBragg}{k}

\newcommand{\tauBragg}{\tau}

\newcommand{\ascritExp}{a_\textrm{s}^{*}}

\newcommand{\DSF}{S(k,\,\omega)}

\newcommand{\DSFsp}{S(k)}

\newcommand{\Nexc}{N_\textrm{exc}}
\newcommand{\omegaRes}{\omega_\kBragg}
\newcommand{\fexc}{\mathcal{F}}

\newcommand{\omegaResk}{\omega_\kBragg}

\newcommand{\kCFinite}{k_\textrm{c}}

\definecolor{grey}{rgb}{0.5, 0.5, 0.5}

\begin{document}
	%\preprint{}
	
\title{High-energy Bragg scattering measurements of a dipolar supersolid}

\author{D.\,Petter}
\affiliation{Institut f\"ur Experimentalphysik, Universit\"at Innsbruck, Technikerstra{\ss}e 25, 6020 Innsbruck, Austria}
	
\author{A.\,Patscheider}
\affiliation{Institut f\"ur Experimentalphysik, Universit\"at Innsbruck, Technikerstra{\ss}e 25, 6020 Innsbruck, Austria}	
	
\author{G.\,Natale}
\affiliation{Institut f\"ur Experimentalphysik, Universit\"at Innsbruck, Technikerstra{\ss}e 25, 6020 Innsbruck, Austria}
	
\author{M.\,J.\,Mark}
\affiliation{Institut f\"ur Experimentalphysik, Universit\"at Innsbruck, Technikerstra{\ss}e 25, 6020 Innsbruck, Austria}
\affiliation{Institut f\"ur Quantenoptik und Quanteninformation, \"Osterreichische Akademie der Wissenschaften, Technikerstra{\ss}e 21a, 6020 Innsbruck, Austria}		
		
\author{M.\,A.\,Baranov}
\affiliation{Institut f\"ur Quantenoptik und Quanteninformation, \"Osterreichische Akademie der Wissenschaften, Technikerstra{\ss}e 21a, 6020 Innsbruck, Austria}

\author{R.\,v.\,Bijnen}
\affiliation{Institut f\"ur Quantenoptik und Quanteninformation, \"Osterreichische Akademie der Wissenschaften, Technikerstra{\ss}e 21a, 6020 Innsbruck, Austria}
\affiliation{Center for Quantum Physics, University of Innsbruck, Austria}

\author{S.\,M.\,Roccuzzo}
\affiliation{INO-CNR BEC Center and Dipartimento di Fisica,
Universit\`a degli Studi di Trento, 38123 Povo, Italy}
\affiliation{Trento Institute for Fundamental Physics and Applications, INFN, 38123, Trento, Italy}		

\author{A.\,Recati}
\affiliation{INO-CNR BEC Center and Dipartimento di Fisica,
Universit\`a degli Studi di Trento, 38123 Povo, Italy}
\affiliation{Trento Institute for Fundamental Physics and Applications, INFN, 38123, Trento, Italy}	

\author{B.\,Blakie}
\affiliation{The Dodd-Walls Centre for Photonic and Quantum Technologies, University of Otago, Dunedin 9054, New Zealand}
\affiliation{Department of Physics, University of Otago, Dunedin 9016, New Zealand}

\author{D.\,Baillie}
\affiliation{The Dodd-Walls Centre for Photonic and Quantum Technologies, University of Otago, Dunedin 9054, New Zealand}
\affiliation{Department of Physics, University of Otago, Dunedin 9016, New Zealand}

\author{L.\,Chomaz}
\affiliation{Institut f\"ur Experimentalphysik, Universit\"at Innsbruck, Technikerstra{\ss}e 25, 6020 Innsbruck, Austria}	

\author{F.\,Ferlaino}
\affiliation{Institut f\"ur Experimentalphysik, Universit\"at Innsbruck, Technikerstra{\ss}e 25, 6020 Innsbruck, Austria}
\affiliation{Institut f\"ur Quantenoptik und Quanteninformation, \"Osterreichische Akademie der Wissenschaften, Technikerstra{\ss}e 21a, 6020 Innsbruck, Austria}	

\date{\today}

\begin{abstract}
We present an experimental and theoretical study of the high-energy excitation spectra of a dipolar supersolid. Using Bragg spectroscopy, we study the scattering response of the system to a high-energy probe, enabling measurements of the dynamic structure factor. We experimentally observe a continuous reduction of the response when tuning the contact interaction from an ordinary Bose-Einstein condensate to a supersolid state. Yet the observed reduction is faster than the one theoretically predicted by the Bogoliubov-de-Gennes theory. 
Based on an intuitive semi-analytic model and real-time simulations, we primarily attribute such a discrepancy to the out-of-equilibrium phase dynamics, which although not affecting the system global coherence, reduces its response.   
\end{abstract}

\maketitle

The field of quantum gases has moved towards the study and realization of novel quantum states of matter, often showing exotic properties~\cite{Bloch2008}. A recent example, still challenging scientist's intuition, is the long-sought supersolid phase, recently observed in  atom-light-coupled systems \cite{Leonard:2017,Li:2017} and dipolar quantum gases~\cite{Tanzi:2019, Boettcher:2019,Chomaz:2019}. A supersolid state spontaneously develops a density modulation in space, breaking the translation symmetry, and a global phase coherence, breaking the gauge symmetry~\cite{Penrose:1956,Gross:1957,Andreev:1969,Chester:1970}. In a dipolar gas, the supersolid phase (SSP) lives in a narrow interaction-parameter range, sandwiched between an ordinary Bose-Einstein condensate (BEC) and an incoherent array of droplets (ID), showing extreme density-modulation~\cite{Lu2015sds, Wenzel:2017, Ancilotto:2019, Zhang2019,  Youssef2019, Boettcher:2019,Chomaz:2019,  Blakie:2020}.

In a dipolar supersolid, fundamental properties related to its quantum-fluid nature remain to be understood. This includes the relation between condensation and superfluidity, as well as their connection to density
modulation and phase fluctuations across the phase diagram of a dipolar gas. 
In a seminal work~\cite{Leggett}, Leggett asks \textit{how solid is a supersolid}, deriving an upper bound for the superfluid fraction of a stationary supersolid state, which connects to the degree of localization, i.\,e.\,the density modulation. In the recently observed dipolar supersolid~\cite{Tanzi:2019,Boettcher:2019,Chomaz:2019}, the situation
might be more complex because of many-body out-of equilibrium
phenomena. Indeed, the macroscopic phase of a supersolid might dynamically develop variations in space, caused e.\,g.\,by the crossing of the BEC-SSP phase transition, or by thermal and quantum fluctuations~\cite{Wenzel:2017, Tanzi:2019,Boettcher:2019,Guo:2019}. Such effects could impact the superfluid properties of the system, going beyond Leggett's original prediction.

Local phase variations are typically not readily accessible in experiments. However, the study of the dynamical response of a physical system to a high-energy scattering probe has proven to contain key information on the state properties.
Such powerful scattering protocols have been widely used across different physical disciplines, ranging from high-energy~\cite{Taylor1991:DIS, Kendall:1991:DIS, Friedman1991:DIS, Breidenbach:1969} to condensed-matter physics~\cite{Griffin:1993, GIACOVAZZO:1992}. For instance, scattering of fast neutrons from superfluid liquid helium has enabled the first measurement of the condensate fraction in a strongly interacting many-body system~\cite{Sosnick:1989}.  In the realm of ultracold quantum gases, a similar concept has been employed to reveal e.\,g.\,beyond-mean-field effects, to measure quantum depletion and coherent fractions, or Tan's universal contact parameter
~\cite{Richard:2003,Esslinger:2004,Papp:2008, Vale:2010, Fabbri:2011, Vale:2013,Lopes:2017}.

In this Letter, we experimentally study the dynamical response of a dipolar supersolid to a high-energy scattering probe by performing two-photon Bragg excitation in the free-particle-excitation regime (high energy and high momentum).  We observe that the system response strongly reduces in the supersolid regime before vanishing in the ID phase.
By benchmarking our data with theoretical models, we identify the role of the density-modulation contrast and the phase variations in the observed response. Our study reveals the importance of the coherent phase dynamics induced by the crossing of the BEC-to-supersolid phase transition.

The dynamical response of an interacting many-body system to a weak scattering probe can be described within the linear-response theory. An essential quantity is the dynamic structure factor (DSF),  $S(\boldsymbol{k},\omega)$, which characterizes the density response of a system to a scattering probe of momentum, $\hbar\boldsymbol{k}$, and energy, $\hbar \omega$~\cite{Pitaevskii:2016}.
For weak interatomic interactions, the DSF can be directly related to the excitation spectrum via the Bogoliubov amplitudes, $u_j$ and $v_j$, describing the excitation mode $j$ of energy $\hbar \omega_j$. It reads
\begin{multline}
    S(\boldsymbol{k},\omega)=\sum_j\bigg|\int d\boldsymbol{r}\left(u_j^*\boldsymbol{(r)}+v_j^*\boldsymbol{(r)}\right)\textrm{e}^{i\boldsymbol{kr}}\psi_0({\boldsymbol r})\bigg|^2\times\\
    \times\delta(\hbar\omega-\hbar\omega_j),
    \label{eq1:dsf}
\end{multline}
where we neglect the creation of multiple excitations. Here, $\psi_0$ is the system's macroscopic ground-state wave function and $\delta(\cdot)$ is the Dirac delta function.

Equation~\eqref{eq1:dsf} gives different information depending on the momentum and energy ranges~\cite{Pitaevskii:2016}: For low-$\boldsymbol{k}$ transfer, $S(\boldsymbol{k},\omega)$ is sensitive to the system's collective response, whereas, in the high-$\boldsymbol{k}$ and high-energy regime, the DSF informs about the momentum distribution of the system, $\tilde n(\boldsymbol{k})$.
We explore the latter regime for our supersolid state, focusing on the response along the density-modulated direction, $y$, with $\boldsymbol{k}=(0,k_y,0)$. 
In the regime of free-particle excitations ($u_j\rightarrow e^{ik_jy}$, $v_j\rightarrow0$, $\omega_j \rightarrow \hbar k_j^2/2m$ with $m$ the atomic mass), the impulse approximation becomes valid and we find~\cite{Hohenberg:1966,Zambelli2000,Pitaevskii:2016,Hofmann:2017,Chomaz:2020Model}
\begin{equation}
    S(k_y,\omega)=\sum_j \tilde n(0,k_y-k_j,0)\,\delta(\hbar\omega-\hbar\omega_j).
    \label{eq1:dsfIA}
\end{equation}
On resonance, $\omega=\omega_j$ and $k_y=k_j$, the DSF becomes ${S(k_j) \equiv S(k_j,\omega_j)~\propto~\tilde n(\boldsymbol{k}=0)}$ and is uniquely determined by the system's momentum distribution at ${\boldsymbol{k}=0}$.

\begin{figure}[ht]
	\includegraphics[width=\columnwidth]{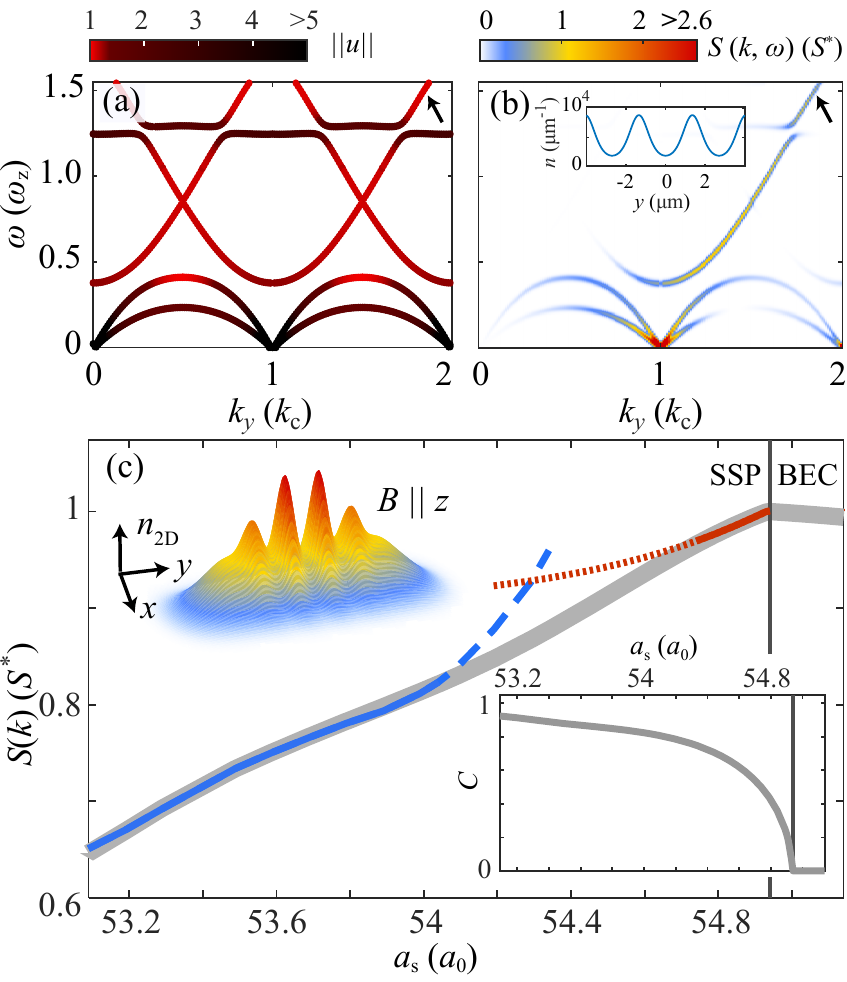}
	\caption {
	(a) Axial excitation spectrum of the transversely symmetric modes and (b) corresponding DSF of an infinitely elongated dipolar supersolid at $\as=51\,a_0$ in a harmonic trap with  $\omega_{x,y,z}=2\pi\times(250,0,160)$\,Hz. The color maps correspond to $\left\lVert u\right\rVert$ and $S(k,\omega)$, respectively. The inset shows the integrated axial density profile $n(y)$ of the ground state with mean density $4.7\times10^3 \upmu\textrm{m}^{-1}$.
	(c) $\DSFsp$ for the 3D-trapped system with $\omega_{x,y,z}=2\pi\times(250,31,160)$\,Hz. $\DSFsp$ is calculated at $k=4.2\,\upmu\mathrm{m}^{-1}\approx1.8\,\kCFinite$ (grey line) and normalized by its value at the BEC-SSP phase transition, $S^*$. The atom number is varied with $\as$ to match the experimental conditions~\cite{supmat}. The red (blue) line shows the result from the SIA (DAA). (upper inset) Integrated density profile of the ground state at $\as=54.49\,a_0$ and $N=5\times10^4$ atoms.
	(lower inset) Evolution of the ground state's central contrast $C$. For the infinite (3D-trapped) case,  $\kCFinite=2.3 (2.4)\,\upmu\mathrm{m}^{-1}$.
	}
	 \label{fig1:DSF} 
\end{figure}

To identify the free-particle regime in our system, we briefly review the basic properties of the excitation spectrum of a supersolid, considering the simpler case of an infinitely elongated cigar-shaped trap.  A more extensive description, including calculations for the full three-dimensional (3D) confined case, which we use hereafter for comparison with the experiments, can be found in Refs.\,\cite{Natale:2019,Tanzi2:2019,Guo:2019, Hertkorn:2019, Roccuzzo19, supmat}. 
As shown in Fig.\,\ref{fig1:DSF}\,(a), the supersolid spectrum exhibits a periodic structure in momentum space with a period given by the reciprocal lattice vector $\kCFinite$. The state develops a density modulation along the axial direction with wavelength $2\pi/\kCFinite$ (see inset in Fig.\,\ref{fig1:DSF}\,(b)). The two lowest branches correspond to the superfluid and crystal branches, respectively~\cite{Saccani:2012}. They have been recently investigated in experiments~\cite{Natale:2019,Tanzi2:2019,Guo:2019}.
The upper branch, appearing at high energy and showing a gapped parabolic dispersion, is the one of interest here for its free-particle character. In addition, the flat band at $\omega\approx 1.25\,\omega_z$ corresponds to a single-droplet excitation (mainly transverse breathing modes), which couples to the parabolic branch, opening small energy gaps.
Figure~\ref{fig1:DSF}\,(b) shows the corresponding DSF values. Interestingly, we observe that the DSF does not reflect the periodicity of the energy spectrum, and e.\,g.\, for the upper branch it shows large values at the momenta that continuously connect to the free-particle excitations in the ordinary BEC~\cite{supmat}.

In a contact-interacting BEC, the inverse healing length provides the scale of the crossover from the collective to the free-particle character of the excitations~\cite{Pitaevskii:2016}. This notion can not be simply exported to the case of dipolar gases because of the momentum dependence of the dipole-dipole interaction. We thus follow the definition based on the Bogoliubov-de-Gennes (BdG) theory.
A free-particle excitation is an elementary excitation whose wave function is well approximated by a plane wave. This is typically justified for excitations of high enough energy and  single-particle character (${\left\lVert u_j\right\rVert=\int |u_j(\boldsymbol{r})|^2d\boldsymbol{r}=1}$ and $\left\lVert v_j\right\rVert = 0$, see color map Fig.\,\ref{fig1:DSF}\,(a))~\cite{Ronen:2006,Pitaevskii:2016}.
For our parameters, we find that the density profile of the mode, $|u_j(\boldsymbol{r})|^2$, shows a plane-wave character for $\hbar\omega\gtrsim0.6\,\hbar\omega_z$.

To quantitatively compare theory with experiments, we perform similar ground-state and BdG-spectrum calculations for the 3D-trapped case~\cite{Natale:2019}, and extract $\DSFsp$~\cite{supmat}.  
As shown in Fig.\,\ref{fig1:DSF}\,(c), in the free particle regime ($k \approx 1.8\kCFinite$), $\DSFsp$ starts to decrease when entering the SSP and further reduces by lowering $\as$. Simultaneously, the ground-state density develops a spatial modulation (upper inset), whose contrast $C$ rapidly increases (lower inset). Note that $C$ evolves faster with $\as$ than $\DSFsp$. For instance, at $\as=53\,a_0$, $C \approx 1$, whereas $\DSFsp$ reduces only by about $35\,\%$. Here, ${C=(n_\mathrm{max}-n_\mathrm{min})/(n_\mathrm{max}+n_\mathrm{min})}$ with $n_\mathrm{max}$ ($n_\mathrm{min}$) being the central maximum (minimum) of the integrated density~\cite{supmat}.

To gain an intuitive understanding of the density-response reduction, we develop a 1D model~\cite{Chomaz:2020Model}. Using two different wavefunction ansatzes, we evaluate $\DSFsp$ in the weak and strong density-modulation regimes. As discussed in Refs.\,~\cite{Blakie:2020,Lu2015sds,Nestor:2008}, for weakly modulated supersolids, with ${C} \ll 1$, the ground-state wave function can be approximated by a fully coherent sine-modulated function on top of a uniform background. At leading order in $C$, it reads $\psi(y)=\sqrt{n}\left(1+C\sin(\kCFinite y)/2\right)$, with $n$ the mean density.
%Plugging
Applying this sine ansatz (SIA) in Eq.\,\eqref{eq1:dsfIA}, we find $\DSFsp \propto n(1-C^2/8)$. 
This result shows that an increasing contrast directly causes a suppression of the DSF.
We find a similar $C$-dependence for the superfluid fraction derived from Leggett's formula~\cite{Leggett}, $f_{ \rm SF}= 1-C^2/2$. 
Therefore, in the weakly modulated regime, the reduction of the high-energy scattering response connects to the reduction of the superfluid fraction.
We benchmark our SIA results with the BdG calculations by evaluating $C$ from the full GPE solution~\cite{supmat}. As shown in Fig.\,\ref{fig1:DSF}\,(c), despite its simplicity, the SIA scaling reproduces very well the full numerics up to $C\lesssim40\,\%$. For larger $C$, as expected, the model breaks down.

For large $C$, we employ a droplet-array ansatz (DAA), describing the system
as an array of $N_D$ droplets, $\psi(y) = \sum_{j=1}^{N_D}\chi(y-jd)e^{i\theta_j}$~\cite{Wenzel:2017,Chomaz:2019}. Each droplet is described by a Gaussian function, $\chi(y)$, of size $\sigma$, separated by a distance $d>\sigma$ from its neighbours. Each droplet is allowed to have an independent, yet uniform, phase $\theta_j$.
Within the DAA, the DSF shows the proportionality $\DSFsp\propto n\left|\frac{1}{N_\mathrm{D}}\sum_{j=1}^{N_D}e^{i\theta_j}\right|^2\sigma/d$. It
decreases with both the density overlap between droplets, set by $\sigma/d$, and the phase variance along the array. The latter effect is not included in the BdG theory, which describes a state possessing a uniform phase.
To benchmark the DAA results with the BdG calculations, we thus set $\theta_j=0$ for all $j$~\cite{supmat}. We find a very good agreement for $C>80\,\%$.

\begin{figure}[t]
	\includegraphics[width=\columnwidth]{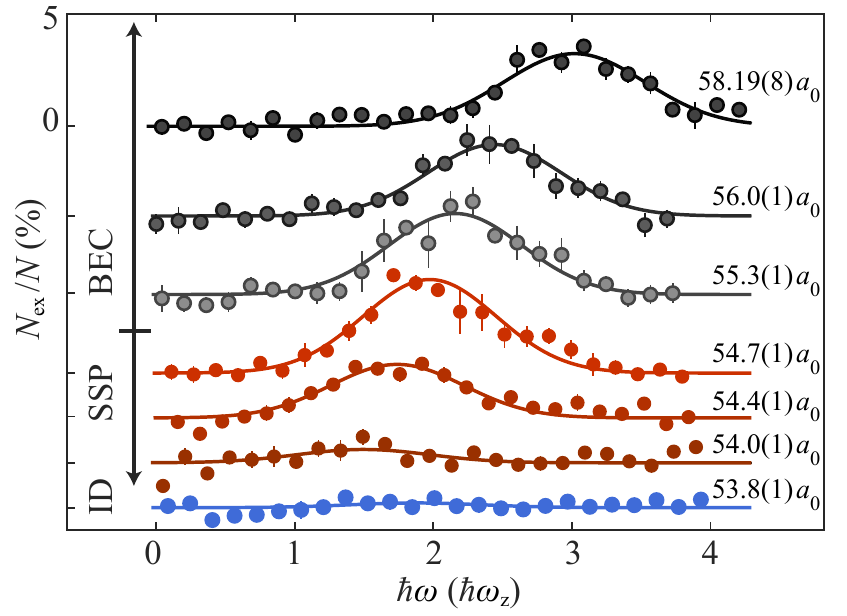}
	\caption {
	Fraction of Bragg-excited atoms as a function of $\omega$ for various $\as$ across the BEC-SSP-ID regimes (see labels). The spectra are vertically offset for visibility. Here and throughout the Letter, the error bars correspond to one standard error. Solid lines show the Gaussian fits to the data.
	}
	 \label{fig2:FiniteSystem} 
\end{figure}

In the experiments, we access the density response of a supersolid by performing high-energy Bragg scattering on a $^{166}\mathrm{Er}$ dipolar quantum gas, confined in an axially elongated harmonic trap.
A transverse homogeneous magnetic field orientates the atomic dipoles and sets $\as$~\cite{Chomaz:2019}. We initially prepare the system in the ordinary BEC phase, and enter the SSP via interaction tuning by linearly lowering $\as$ below a critical value, $\ascritExp$, for which the BEC-SSP phase transition occurs. Similar to previous experiments~\cite{Chomaz:2019, Natale:2019}, $\ascritExp$ is extracted with an interferometric technique. For the present trap and atom numbers, $N$, we measure $\ascritExp=54.94_{-13}^{+28}\,a_0$; see~\cite{supmat}.

For the Bragg excitation, we project on the atoms an optical lattice potential of constant depth $V$ for a duration $\tau=7\,$ms. The lattice has a constant wave vector $\kBragg=4.2(3)\upmu\textrm{m}^{-1}$ along $y$ and moves with a variable frequency $\omega$. 
After the Bragg excitation, we measure the integrated momentum distribution, $\tilde{n}(\kBragg_x,\kBragg_y)$, using a time-of-flight expansion of 30\,ms. The number of excited atoms $N_\textrm{exc}$ is extracted in a narrow region of interest around $\kBragg$~\cite{supmat}. For a fixed $\as$, we find a clear resonance in $N_\textrm{exc}/N$ as we vary $\omega$. From a Gaussian fit we extract the resonance peak's amplitude, $\fexc$. From linear response theory, we expect  $\fexc \propto V^2 \tauBragg\DSFsp$~\cite{Blakie:2002}. For the relevant $\as$ range, we have checked the scaling with $\tauBragg$ and $V$~\cite{supmat}.
Figure~\ref{fig2:FiniteSystem} shows examples of the Bragg-excitation spectrum for various $\as$. In the BEC regime until the onset of the SSP, we observe a downward shift of the resonance frequency without a significant change in $\fexc$~\cite{supmat}. In contrast, as we enter into the SSP regime, $\fexc$ undergoes a stark reduction. In the ID regime, the resonance peak completely vanishes.

%%----- Measurements-Theory comparison
\begin{figure}[htb]
	\includegraphics[width=\columnwidth]{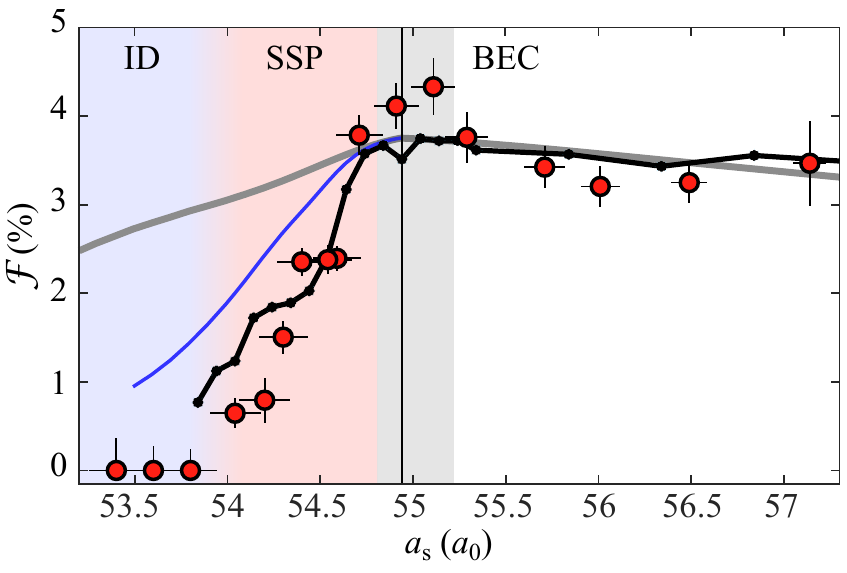}
	\caption {Experimental $\fexc$ (circles) versus $\as$ across the BEC-SSP-ID phases (white, red, blue shadings). For the lowest three $\as$, we do not observe a resonance and plot the standard deviation of the data as an error estimate. Horizontal error bars correspond to uncertainties of the magnetic field~\cite{supmat}. We show the excited fraction expected from the BdG calculations on the corresponding ground states (gray line) and from the RTE simulations (connected dots, one RTE simulation run). In the RTE we cannot extract clear resonances for the lowest $\as$. We furthermore show the rescaled BdG calculations, which include $\Delta\Theta$ obtained from the RTE (blue line). The gray shading corresponds to the uncertainty in $\as$ of the experimental phase transition (vertical line).
	}
	 \label{fig3:ExpTheoryComp} 
\end{figure}

Figure~\ref{fig3:ExpTheoryComp} shows the evolution of $\fexc$ across the BEC-SSP-ID phase diagram. The $\as$-extension of the three phases (see background colors), has been determined from independent measurements of the phase coherence and density modulation of the states~\cite{Chomaz:2019, supmat}.
When reducing $\as$ in the BEC phase, we find a slight increase of $\fexc$. On both sides of the BEC-SSP phase transition, we observe similar $\fexc$ values, indicating a continuous behavior across the transition. Remarkably, as soon as we lower $\as$ further by $\sim$0.5\,$a_0$, $\fexc$ drastically reduces to $\lesssim1\,\%$, which is close to our detection level. Finally, for $\as < 54\,a_0$, we do not observe any resonant response.

We compare the experimental results with our BdG theory for the 3D-trapped gas. While in the BEC regime, experiment and theory show a good agreement, in the SSP they start to substantially deviate from each other. The data shows a much faster reduction of the system's response than the one predicted from the BdG theory. This result suggests that an important ingredient is missing in the theory. Our DAA model provides a first intuitive explanation: It suggests that the presence of phase variations across the system can yield a reduction of the DSF. We envision two sources of phase variations. First, quantum and thermal fluctuations, which are expected to dominate in the ID regime, yield phase patterns varying from shot to shot. Second, coherent dynamics, as e.\,g. induced by the crossing of the BEC-SSP phase transition, leading to reproducible phase patterns.
Neither phenomena are accounted for in the BdG calculations.

To investigate these effects, we simulate the system real-time evolution (RTE)~\cite{Chomaz:2018}. Our calculations reproduce the experimental sequence and include the linear ramp in $\as$, the Bragg excitation, the three-body losses, and an initial population of BdG modes from quantum and thermal noises~\cite{supmat}. 
From the simulated momentum distributions, we extract the excited fractions, following the same procedure as for the experimental ones. As shown in Fig.\,\ref{fig3:ExpTheoryComp}, contrary to the BdG results, the 
RTE simulations describe remarkably well the data both in the BEC and SSP phase.

\begin{figure}[ht]
	\includegraphics[width=\columnwidth]{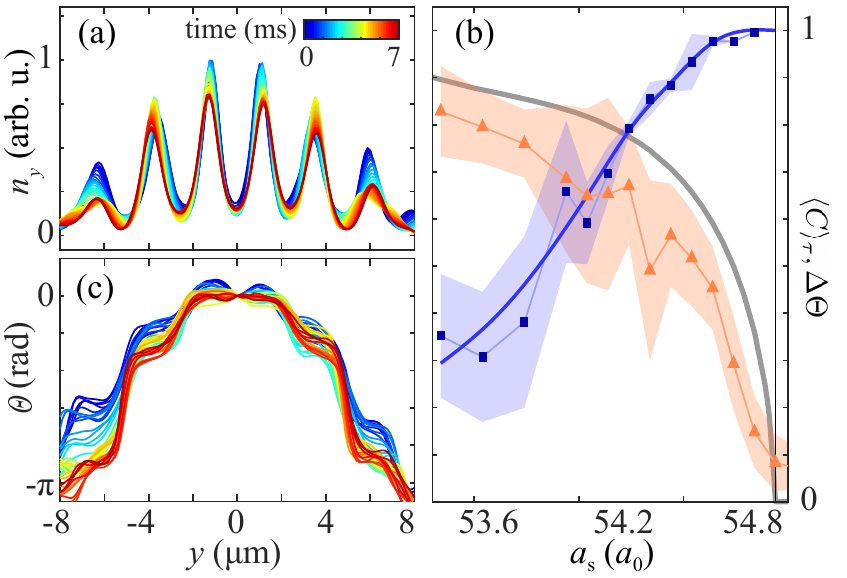}
	\caption {
	RTE simulations without Bragg excitation. (a) Time evolution of the integrated in-situ density of the wave function for $\as=54.04\,a_0$. (b) $\langle C\rangle_\tauBragg$ (triangles) and $\Delta\Theta$ (squares) versus $\as$. The grey line corresponds to the central contrast obtained from the ground-state theory. The solid blue line is a smooth interpolation of $\Delta\Theta$, fixed to unity at the phase transition point. The shadings give the standard deviation obtained from 5 simulation runs. The vertical line corresponds to the phase transition point. (c) Phase-cuts corresponding to the simulation shown in (a).
    }
	 \label{fig4:timeEvolRTE} 
\end{figure}

To highlight the role of the contrast and phase variations, we perform RTE simulations without the Bragg excitation for different holding times. 
As shown in Fig.\,\ref{fig4:timeEvolRTE}\,(a), the density profiles $n(y)$ exhibit only a slight reduction of the contrast with time due to atom loss. As expected, the calculated $\langle C\rangle_\tauBragg$,  time-averaged over the Bragg scattering duration, increase with decreasing $\as$. However, for each $\as$, we observe a 10-30\,\% lower contrast than the one extracted from the ground-state calculations.
Since a reduced contrast would mean an increase in $\fexc$, we deduce that the varying contrast can not explain the mismatch between the BdG theory and both the experimental and RTE results; see Fig.\,\ref{fig3:ExpTheoryComp}.

We now study the phase variations and its dynamics. The RTE calculations reveals that the phase of the wave function, $\theta(y)$, develops a non-uniform profile. For instance at $\as=54.04\,a_0$, $\theta(y)$ exhibits a stair-like profile with fairly constant values within the density peaks and discrete phase steps in between them; see Fig.\,\ref{fig4:timeEvolRTE}\,(c). This behaviour suggests that each density peak acquires an independent phase, despite their density links. We also observe that the phase patterns is fairly reproducible between simulation runs and mainly induced by the coherent dynamics arising by the crossing of the phase transition~\cite{supmat}. 

Following the DAA model, phase variations are expected to reduce $\DSFsp$ by a factor 
$\Delta\Theta\approx|\frac{1}{N_\mathrm{D}}\sum_{j=1}^{N_D}\langle e^{i\theta_j}\rangle_\tauBragg|^2$~\cite{Pitaevskii:2016, supmat}. 
As shown in Fig.\,\ref{fig4:timeEvolRTE}\,(b), $\Delta\Theta$ is almost unity close to the BEC-SSP phase transition and significantly drops when lowering $\as$ towards the ID regime, where  starts to flatten. The standard deviation of $\Delta\Theta$ relates to the shot-to-shot reproducibility of the phase pattern.  In the SSP, the deviation remains small, confirming that the phase variations originate from coherent dynamics. In contrast, the deviation increases when reaching the ID regime, highlighting the increasing effects of fluctuations. We empirically account for the effect of phase variations in the BdG theory by scaling the DSF with $\Delta\Theta$ over the whole SSP-ID regimes. As shown in Fig.\,\ref{fig3:ExpTheoryComp}, this simple inclusion of $\Delta\Theta$ demonstrates the pronounced impact of the coherent phase variations for the experimentally observed response.

In conclusion, we demonstrate that the supersolid states, when created via a dynamical tuning of the interactions, develop important phase variations across the system, which have to be taken into account to understand the system behavior. Those phase variations occur even in presence of sufficiently strong density links between the droplets.
Our work provides first steps to a more complete vision of the dipolar supersolid, including out-of-equilibrium phenomena, and opens the door for future exploration of critical phenomena induced by the dynamical crossing of the BEC-SSP phase transition~\cite{Kibble:1976,Zurek:1985,Eisert:2015}.

\begin{acknowledgments}
 
We thank S.\,Stringari for insightful discussions and B. Yang for the careful reading of the manuscript. Part of the computational results presented have been achieved using the HPC infrastructure LEO of the University of Innsbruck. This work is financially supported through an ERC Consolidator Grant (RARE, no.\,681432), a DFG/FWF (FOR 2247/PI2790) and a joint-project grant from the FWF (I 4426, RSF/Russland 2019). L.\,C.\,acknowledges the support of the FWF via the Elise Richter Fellowship number V792. A.\,R.\,and S.\,M.\,R.\,acknowledge support from Provincia Autonoma di Trento and the Q@TN initiative. We also acknowledge the Innsbruck Laser Core Facility, financed by the Austrian Federal Ministry of Science, Research and Economy.
\end{acknowledgments}

* Correspondence and requests for materials
should be addressed to Francesca.Ferlaino@uibk.ac.at.

\bibliography{BibBraggSupSol}

%%%%%%%%%%%%%%%%%%%%%%%%%%%%%%%%%%%%%%%%%%%%%%%%%%%%
\clearpage
\appendix

\renewcommand\thefigure{\thesection S\arabic{figure}}   
\setcounter{figure}{0}   

\section{Supplemental Material}

\subsection{A.\,Preparation of the BECs}
We prepare a BEC of $\Er$ by loading about $3\times10^6$ thermal atoms into a crossed optical dipole trap (ODT) and subsequent evaporative cooling, see Ref.\,\cite{Chomaz:2016, Chomaz:2019}. During the evaporative cooling, a homogeneous magnetic field of 1.9\,G is present to ensure high enough rethermalization rates to obtain ultracold temperatures. After the evaporation, we adiabatically modify the corresponding ODTs laser powers and beam waists, to shape the confining potential  $V_\mathrm{trap}(\textbf{r})=m(\omega_x^2x^2+\omega_y^2y^2+\omega_z^2z^2)/2$ to a cigar-shaped geometry with harmonic trapping frequencies {$\omega_{x,y,z} = 2\pi\times[250(1), 31.7(13) ,156(2)]\,$Hz}. Consecutively, the magnetic field is lowered to a value corresponding to $64.9\,a_0$. After this preparation procedure, we obtain a BEC with a total atom number of $1.2\times10^5$ atoms and a condensed fraction of $70\,\%$. The temperature of $95(5)\,$nK is obtained from time-of-flight (ToF) expansion measurements.

To enter the BEC-SSP-ID regimes, we lower down $\as$ linearly in 20\,ms to the corresponding values given in the main manuscript. We then let the system equilibrate for 10\,ms and consecutively apply a Bragg pulse of 7\,ms duration. In order to access the momentum distribution of our atomic cloud, we perform a ToF expansion, by abruptly switching off all trapping potentials directly after the Bragg excitation. After 30\,ms of free expansion, we take an absorption image of the cloud along the dipole direction.
We note that, due to residual magnetic field drifts in the experiment (estimated to be $\pm2\,$mG), the uncertainty on $\as$, during the Bragg pulse, ranges between $\pm0.1\,a_0$ and $\pm0.2\,a_0$, increasing for lower $\as$. This uncertainty is represented by the corresponding error bars on our data in the main manuscript.

\subsection{B.\,Determination of experimental BEC-SSP phase transition}
\begin{figure}[ht]
	\includegraphics[width=\columnwidth]{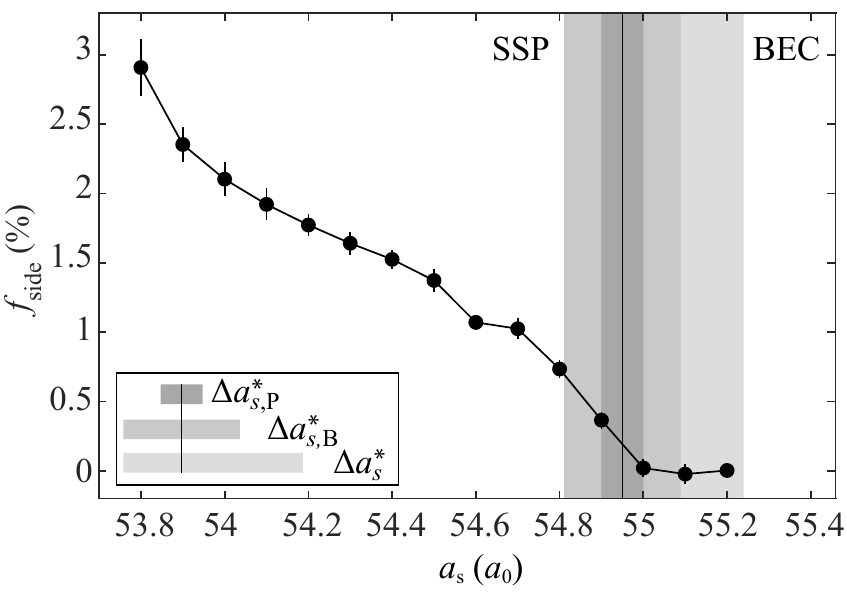}
	\caption {Fraction of atoms in one side peak of the atomic cloud's interference pattern across the BEC-SSP phase transition. Error bars denote one standard error obtained from about 30 measurements. The vertical line shows the obtained phase transition point. The different grey shadings correspond to the different uncertainties that are taken into account to obtain the total uncertainty, $\Delta\ascritExp$, of $\ascritExp$ (see text).}
	\label{figS2:ExpPhaseTrans}
\end{figure}
In order to determine $\ascritExp$ for our experimental parameters, we perform a time-of-flight expansion of the system in the BEC or SSP regime after the equilibration time. Here, no Bragg pulse is applied. We find either an expanded ordinary BEC or an interference pattern of the expanded supersolid, where a part of the atoms appear in two side peaks around $k_y\approx\pm k_\mathrm{c}$. The atom number in these two side peaks is directly related to the modulation contrast of the in-situ cloud~\cite{Chomaz:2019,Tanzi:2019, Boettcher:2019}.
We measure the fraction of atoms in a single side peak, $f_\mathrm{side}$, and monitor its evolution versus $\as$; see Fig.\,\ref{figS2:ExpPhaseTrans}. In the BEC regime, where no density modulation is present, we find  $f_\mathrm{side} = 0$ down to $a_\textrm{s,1}=55.00\,a_0$. After crossing the BEC-SSP phase transition, we observe $f_\mathrm{side} > 0$ for $\as\leq54.88\,a_0=a_\textrm{s,2}$, which increases with lower $\as$.
Taking the mean, $(a_\textrm{s,1}-a_\textrm{s,2})/2$, we find $\ascritExp=54.94\,a_0$ with an uncertainty of $\Delta a_{s,\mathrm{P}}^*=0.05\,a_0$, coming from our resolution in $\as$; see shadings in Fig.\,\ref{figS2:ExpPhaseTrans}. We include a magnetic field uncertainty corresponding to 2\,mG ($\pm0.12\,a_0$ at $\ascritExp$), which increases the uncertainty to $\Delta a_{s,\mathrm{B}}^*=\pm0.13\,a_0$. Furthermore, we take a finite resolution of $f_\mathrm{side}\approx 0.2\,\%$ into account and obtain the final estimate of the critical point of the BEC-SSP phase transition $\ascritExp=54.94_{-13}^{+28}\,a_0$.

\subsection{C.\,Transition from SSP to ID}
\begin{figure}[hbt]
	\includegraphics[width=\columnwidth]{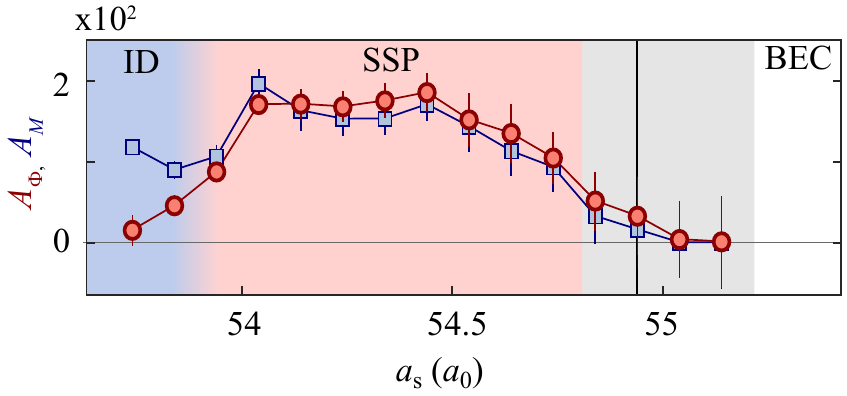}
	\caption {Amplitudes $A_\phi$ (red circles) $A_\mathrm{M}$ (blue squares) versus $\as$. Error bars denote the standard error from about 30 experimental realizations~\cite{Chomaz:2019}.
	Non-zero values of $A_\mathrm{M}$ enable us to identify modulated states and confirms the BEC-SSP transition point (vertical line, gray shaded area). The SSP is identified by $ A_\mathrm{M}\approx A_\phi>0$ and extends down to $\as=53.9\,a_0$. For lower $\as$ an ID state is observed ({$A_\phi<A_\mathrm{M}>0$}).}
	\label{figS6:SSPIDTrans}
\end{figure}
We use the same analysis of $A_\phi$ and $A_\mathrm{M}$ as in Ref.~\cite{Chomaz:2019} to distinguish in the experiment the SSP and the ID regime. In brief, $A_\phi$ relates to a reproducible interference pattern in time of flight and thus reveals a coherent and modulated state. $A_\mathrm{M}$ relates only to the presence of an in-situ density modulation (structures in the ToF images). By combining both observables, one can distinguish the SSP ($A_\phi\approx A_\mathrm{M}>0$), the ID ($A_\phi<A_\mathrm{M}$) and the ordinary BEC regimes ($A_\phi= A_\mathrm{M}=0$) in the experiment, see Fig.\,\ref{figS6:SSPIDTrans}. We find that for $\as<53.9\,a_0$ the system is in the ID regime. The measurements are performed directly after the equilibration time and without a Bragg excitation (same timings as in Sec.\,B).
We note that the ratio $A_\phi/A_M$ is mostly sensitive to phase \textit{fluctuations}, which lead to different interference patterns in different experimental runs and is insensitive to reproducible phase variations in the system. The latter could affect the structure of the interference patterns, yet in a reproducible way. Therefore, $A_\phi/A_M$ is an observable that is adapted to describe the coherence of the system but does not measure the phase variations, investigated in the main manuscript, which are induced by diabatic dynamics.

\subsection{D.\,Calibration of atom loss and atom number for BdG theory}
\begin{figure}[ht]
	\includegraphics[width=\columnwidth]{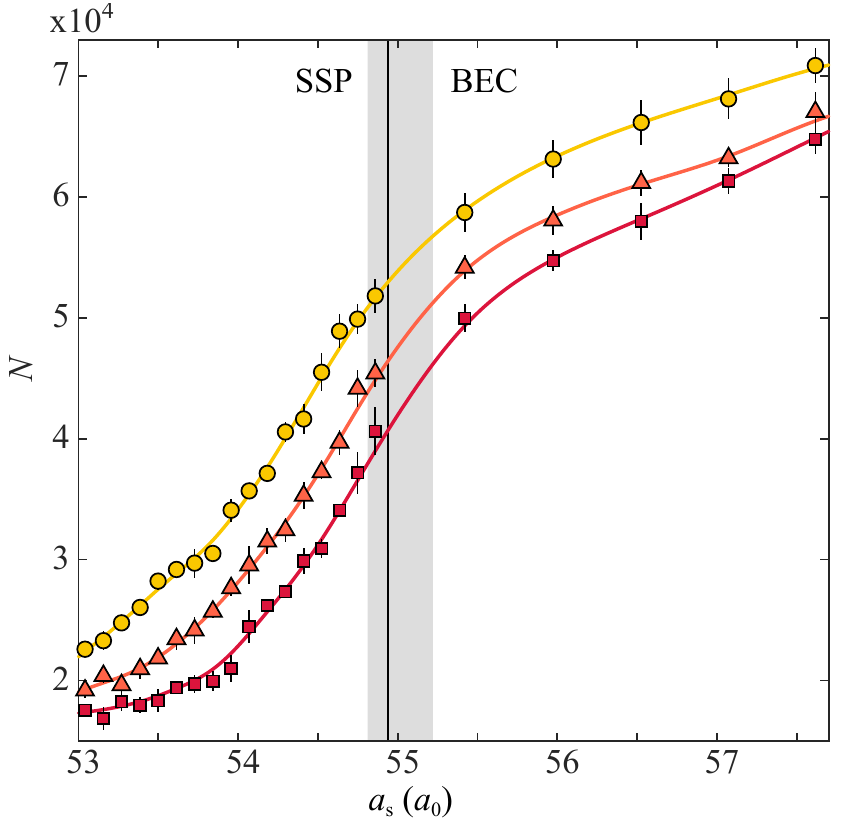}
	\caption {Atom number in the BEC versus $\as$ for different times in the experiment, corresponding to the beginning (circles), middle (triangles) and end (squares) of the Bragg pulse. Error bars denote the standard error from 5 measurements. The lines are spline interpolations to the corresponding data. The measurements were obtained without applying a Bragg pulse. The vertical line indicates the measured phase transition point.}
	\label{figS3:atomNumberVSHoldTime}
\end{figure}
Due to three-body recombination losses, the atom number, $N$, in the condensed part is decreasing during the 7\,ms of the Bragg pulse by typically 10-30\,\%. Therefore, the atom number in the experiment varies with $\as$, which we include in our BdG theory of the three-dimensionally trapped system. We note that we do not observe additional atom loss due to the presence of the Bragg excitation, as the wavelength of the used laser light is far enough detuned from any atomic resonance (see Sec.\,E).

To extract $N$ we perform an additional set of measurements in which we do not apply a Bragg pulse and, after a given hold time in trap, take absorption images after a 30 ms ToF expansion. From these images, the thermal component is fitted by an isotropic 2D Gaussian function and subtracted. A final numerical integration over the image yields $N$ without the need of an additional fitting of the condensed part itself.
In Fig.\,\ref{figS3:atomNumberVSHoldTime}, we show $N$ across the phase diagram for different timings in the experiment, corresponding to the beginning, the middle and the end of the Bragg pulse. Each timing is interpolated with a spline fit. The fitted values of the intermediate timing (orange line in Fig.\,\ref{figS3:atomNumberVSHoldTime}) is used as the atom number in our BdG theory.

\subsection{E. Bragg spectroscopy}

The Bragg excitation beams are realized holographically with a digital-micromirror device (DMD), as detailed in Ref.\,\cite{Petter:2019}. In short, the setup uses a near-resonant laser light, red-detuned by 71(1)\,GHz from the 401\,nm transition of $\Er$.
These two Bragg beams interfere under an angle on the atoms' position, giving rise to an interference pattern. In our setup this angle can be tuned, but for this current work we keep it fixed to obtain an interference pattern with a wave vector $\kBragg=4.2(3)\,\upmu\mathrm{m}^{-1}$ along $y$. The value and uncertainty on $\kBragg$ is deduced from offline measurements of the angle between the two Bragg beams.
To excite the system, the Bragg scattering needs to supply energy, $\hbar\omega$, which is introduced with a frequency difference, $\omega$, between the two Bragg beams. Here, we use a sequence of holographic gratings that is uploaded on the DMD and continuously shifts the phase of one beam in 9\, steps from 0 to $2\pi$. Depending on the frame-rate of the uploaded sequence, we can vary $\omega$ from 0\,Hz to 1000\,Hz.

To calibrate the depth $V$ of our Bragg potential, we perform Kapitza-Dirac-diffraction measurements~\cite{Gould1986}. For these measurements, we tune the laser light closer to the atomic transition ($20.6\,$GHz) and use the maximally available power for our Bragg beams. By doing so, we achieve a maximum optical depth of ${V_\mathrm{max}/h=430(50)}$\,Hz, corresponding to $3.2(4)\,E_\mathrm{rec}$, where the recoil energy $E_\mathrm{rec}=\hbar^2(k/2)^2/(2m)=135\,$Hz.
In order to extract the potential depth $V$ of our Bragg scattering probe, we rescale this calibrated value according to the corresponding laser light detuning and power used for the Bragg scattering~\cite{Bloch2008}. We obtain $V=42(5)\,$Hz, which is well inside the linear scattering regime (see Sec.\,F and Fig.\,\ref{figS5:ExcFracVSTau}). The exact calibration of the potential depth does not include systematic effects, as for example inhomogeneities of $V$ cross the atomic cloud or in-trap dynamics of the atoms during the Kapitza-Dirac-diffraction measurement. Nevertheless, we note that the estimation of the linear response regime is insensitive to the exact calibration of $V$. We furthermore note, that a direct comparison of $\fexc$ from the experiment with the one from the RTE suggests that $V$ needs to be rescaled by about $1.7$.

\begin{figure}[tb]
	\includegraphics[width=\columnwidth]{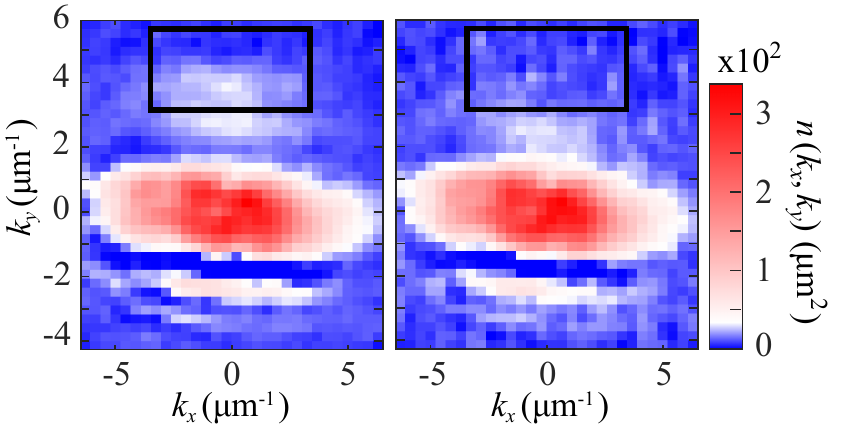}
	\caption {Examples of momentum distributions of a supersolid at $\as=54.59(13)\,a_0$ after an applied Bragg excitation (a) on resonance at $1.7\,\hbar\omega_z$ and (b) off resonance at $3.7\,\hbar\omega_z$. The two side peaks appearing around $k_y\approx\pm2.4\upmu\mathrm{m}^{-1}$ constitute, together with the central peak at $k\approx0\upmu\mathrm{m}^{-1}$, the interference pattern obtained when expanding a supersolid state. The black box indicates the region of interest from which $N_\textrm{exc}$ is extracted. Each image is an average of 15 experimental realizations.}
	 \label{figS14:AbsImgsOnOffRes} 
\end{figure}

Figure~\ref{figS14:AbsImgsOnOffRes} gives examples of our images for a resonant and an off-resonant Bragg scattering frequency. For the resonant case (Fig.~\ref{figS14:AbsImgsOnOffRes}, left panel) we find scattered atoms at a high momentum around $k_y\approx\kBragg$.
As they appear outside of the interference patterns, observed from the unperturbed system, they constitute a clean signal for the analysis of the excited fraction. We count the number of atoms, $N_\mathrm{exc}$, in a region of interest (ROI), indicated by the black boxes in Fig.~\ref{figS14:AbsImgsOnOffRes}. We note that we carefully checked that neither $\fexc$, nor $\omegaRes$, changes within the uncertainties when increasing the ROI size by 30\%.. We measure the total atom number, $N$, for each measurement individually, by performing a similar count on a rectangle of $12\,\upmu\mathrm{m}^{-1}$ by $14\,\upmu\mathrm{m}^{-1}$, covering all condensed and scattered atoms. By measuring $\Nexc /N$ for different excitation frequencies, we obtain a spectroscopy of the Bragg scattering. We note that due to thermal atoms, present in the analyzed region of interest, all Bragg resonances show a small offset, which is extracted from the offset of the Gaussian fit to the resonance and then subtracted.

\begin{figure}[ht]
	\includegraphics[width=\columnwidth]{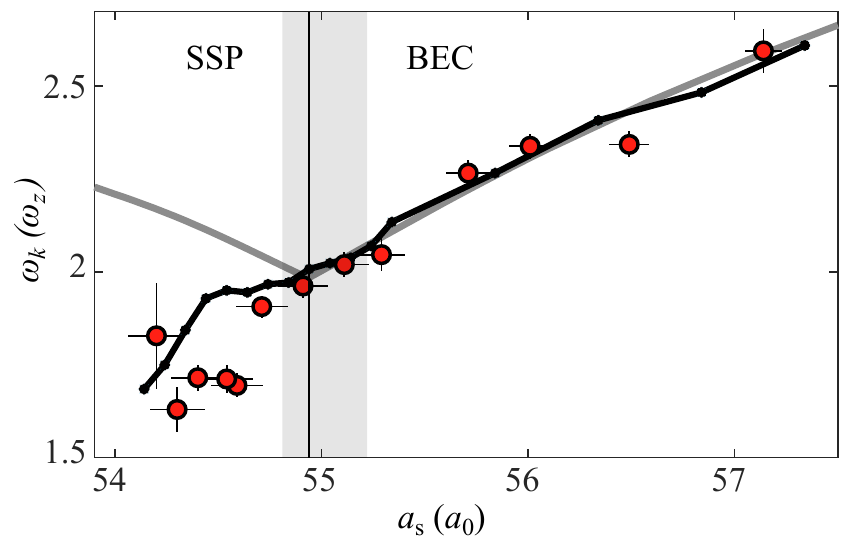}
	\caption {Extracted resonance frequencies, $\omegaResk$, versus $\as$ (circles) and their comparison with the expected resonance position from the BdG theory (gray line) and $\omegaResk$ from the RTE (connected dots). The vertical line indicates the BEC-SSP phase transition point and its shading the uncertainty on $\ascritExp$.}
	 \label{FigS10:resFreqVSas}
\end{figure}

From the Bragg excitation spectra, we extract the resonance peak's amplitude, as discussed in the main manuscript, and a resonance frequency, $\omegaResk$. The latter is shown in Fig.\,\ref{FigS10:resFreqVSas} as a function of $\as$. We observe that $\omegaResk$ decreases monotonically from high to low $\as$, across the BEC-SSP phase transition. This beheviour is consistent with the extracted $\omegaResk$ from the RTE theory. Furthermore, we calculate the expected resonance frequency from the BdG calculations, in which the resonance frequency is increasing again after crossing the BEC-SSP phase transition with lowering $\as$. 
Therefore, the BdG theory predicts a hardening of the measured excitation modes, which is not observed neither in the experiment, nor in the RTE simulations. This qualitative difference might stem from the increased phase variations that develop in the system, but further studies are needed to elucidate this point.

\subsection{F. Variations of the excited fraction with the Bragg pulse duration}

\begin{figure}[h]
	\includegraphics[width=\columnwidth]{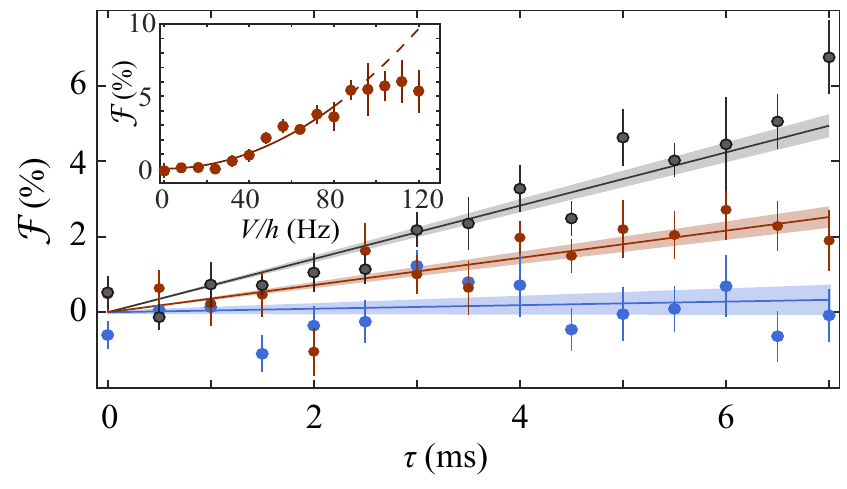}
	\caption {Evolution of $\fexc$ during the Bragg pulse for three exemplary $\as=[55.5(1),\,54.0(1),\,53.3(1)]\,a_0$ (black, red, blue), corresponding to the BEC, SSP, ID regimes, respectively. Each data point corresponds to an average of 5 to 15 measurements and its uncertainty to the standard error. The solid lines correspond to a linear fit and its shading to the fit's 68\,\%-confidence bound.
	The inset shows the evolution of $\fexc$ with the applied potential depth, $V$, of the Bragg pulse for $\tauBragg=7\,$ms for a supersolid state at $\as=54.28(14)\,a_0$. The solid line corresponds to a quadratic fit up to $V=80\,$Hz, the dashed line is the extension of the fitting.}
	 \label{figS5:ExcFracVSTau} 
\end{figure}
In Figure~\ref{figS5:ExcFracVSTau}, we present the measured evolution of $\fexc$ with the Bragg pulse duration from 0 to 7\,ms for a fixed $\omega$. Across the BEC-SSP-ID regimes, we find a linear scaling of $\fexc$ with $\tauBragg$, which is consistent with the expected scaling from BdG theory, $\fexc\propto V^2 \tauBragg\DSFsp$~\cite{Blakie:2002}. Furthermore, we probe the quadratic scaling of $\fexc$ with $V$ in the SSP and find also here an agreement up to $V=80\,$Hz (see inset).

\subsection{G. Evolution of the excitation spectrum with $\as$ for the infinite cigar-shaped gas}

\begin{figure*}[ht]
	\includegraphics[width=\textwidth]{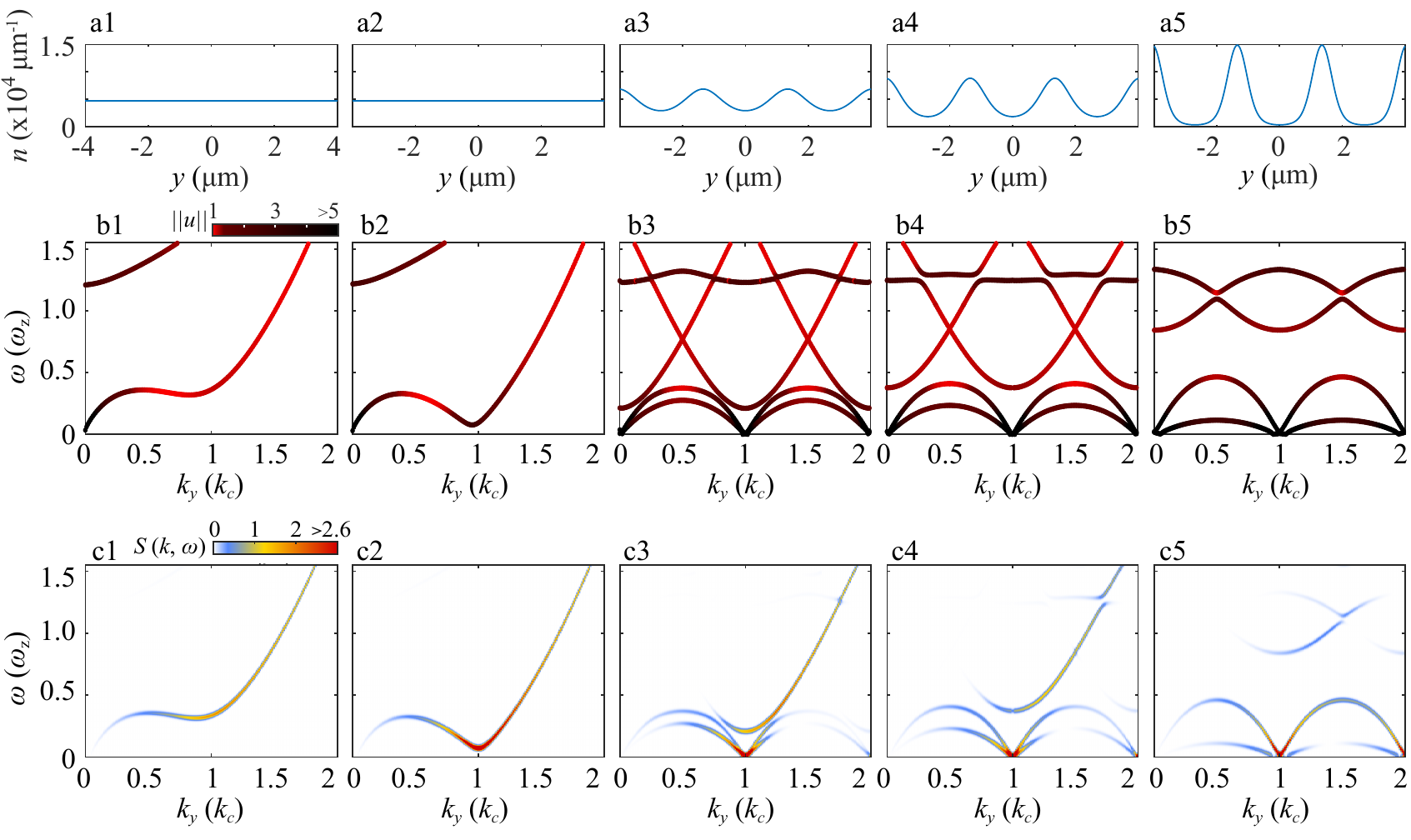}
	\caption {Axial excitation spectra of the infinitely, extended gas across the BEC-SSP-ID regimes in a $\omega_{x,y,z}=2\pi(250,0,160)$\,Hz trap with fixed axial density $n_0=4.7\times10^3 \upmu\textrm{m}^{-1}$. (a1-a5) Integrated density profiles along the unconfined direction, for $\as=(52.00,\,51.40,\,51.25,\,51.00,\,49.75)\,a_0$. (b1-b5) Transverse symmetric modes of the corresponding excitation spectrum colored according to $\left\lVert u\right\rVert$. (c1-c5) The corresponding $\DSF$. For visibility, the DSF is broadened with a Gaussian function.
	}
	\label{figS11:ExcSpecInfTheory}
\end{figure*}
We calculate the excitation spectrum and the dynamic structure factor (DSF) of an infinitely elongated, cigar-shaped dipolar supersolid in Fig.\,1(a,\,b) of the main manuscript.
In Figure\,\ref{figS11:ExcSpecInfTheory} we present the evolution of the excitation spectrum across the BEC-SSP-ID regimes.
Figure\,\ref{figS11:ExcSpecInfTheory}\,(a1-a5) shows the integrated density profiles of the ground state along the unconfined direction for different values of $\as$ and a fixed mean axial density of $4.7\times10^3\,\upmu\mathrm{m}^{-1}$. Figure\,\ref{figS11:ExcSpecInfTheory}\,(b1-b5) shows the corresponding excitation spectrum. At large enough $\as$, the ground state has a uniform density along the unconfined direction (a1,\,a2 - BEC phase) and its excitation spectrum shows the typical phonon-maxon-roton spectrum, first predicted in \cite{Santos:2003, ODell2003}. When decreasing $\as$ below a critical value, the ground state becomes density modulated (a3,\,b3 - SSP phase) with a modulation wave number $k_\textrm{c}=2.3\,\upmu\textrm{m}^{-1}$ close to the BEC's roton momentum (b2), underlying the connection between roton softening and crystallization.
The density modulation has a finite contrast and its value increases when lowering $\as$ further down (a4,\,a5).

When crossing the BEC-SSP phase transition, the excitation spectrum changes dramatically, becoming periodic, with the appearance of two gapless Goldstone branches associated with phase (lower energy branch) and density (higher energy branch) excitations, respectively \cite{Macri:2013, Saccani:2012, Natale:2019}.
In addition to these gapless branches, one observes gapped parabolic branches of excitations with energetic minima at integer multiples of $k_\textrm{c}$. The one branch at $k_y=k_\textrm{c}$ is the one investigated in the main manuscript. For decreasing $\as$, the energy minimum of this parabolic branch increases towards the ID regime [Fig.\,\ref{figS11:ExcSpecInfTheory}\,(b3-b5)].

As described in the main paper, we use the norm of the calculated Bogoliubov amplitude $\left\lVert u_j\right\rVert = \int |u_j(\textbf{r})|^2d\textbf{r}$ to distinguish whether a mode $j$ is a collective excitation or has a single-particle character \cite{Ronen:2006,Pitaevskii:2016}.
Collective excitations feature $\left\lVert u_j\right\rVert\gg1$ whereas single-particle excitations have $\left\lVert u_j\right\rVert\simeq1$.
In Figure\,\ref{figS11:ExcSpecInfTheory}\,(b1-b5), we color each excitation mode according to $\left\lVert u\right\rVert$. We find that lower energy modes, such as the roton mode in the BEC and the Goldstone modes in the SSP have a clear collective nature. The energetically higher modes ($\hbar\omega\gtrsim0.5\,\hbar\omega_z$), of the parabolic branch in the SSP and the $k_y>k_\textrm{c}$-branch in the BEC, have $\left\lVert u_j\right\rVert\simeq1$ across the BEC-SSP phase transition.
We note that the condition $\left\lVert u_j\right\rVert\simeq1$ does not directly identify an excitation mode as a free-particle. To obtain a free-particle excitation, the mode needs to be of single particle character and additionally its energy needs to be mainly given by the kinetic energy. Therefore, free-particle excitations have a wave function that is a plane wave~\cite{Pitaevskii:2016}. We note that for our experimentally relevant energy regime, the probed excitation modes are described well by a plane wave, as shown in Sec.\,I. and Fig.\,\ref{figS4:FreeParticleRegime}\,(b).

From our simulations, we also calculate the DSF. In the BEC phase [Fig.~\ref{figS11:ExcSpecInfTheory}\,(c1,\,c2)] the DSF is dominated by the roton mode at $k_y=k_\textrm{c}$. Moreover, the deeper the roton minimum, the stronger is its response to small density perturbations. We note that, in the BEC phase, this affects also the density response even for momenta higher then the roton momentum. After crossing the phase transition into the SSP, we find that the DSF of the parabolic branch [Fig.~\ref{figS11:ExcSpecInfTheory}\,(c3)] smoothly connects to the free-particle branch of the BEC phase. For decreasing $\as$, the DSF of the free particle branch becomes smaller [Fig.~\ref{figS11:ExcSpecInfTheory}\,(c4,\,c5)].

\subsection{H. BdG theory for the three-dimensional trapped gas}

\begin{figure}[htb]
	\includegraphics[width=\columnwidth]{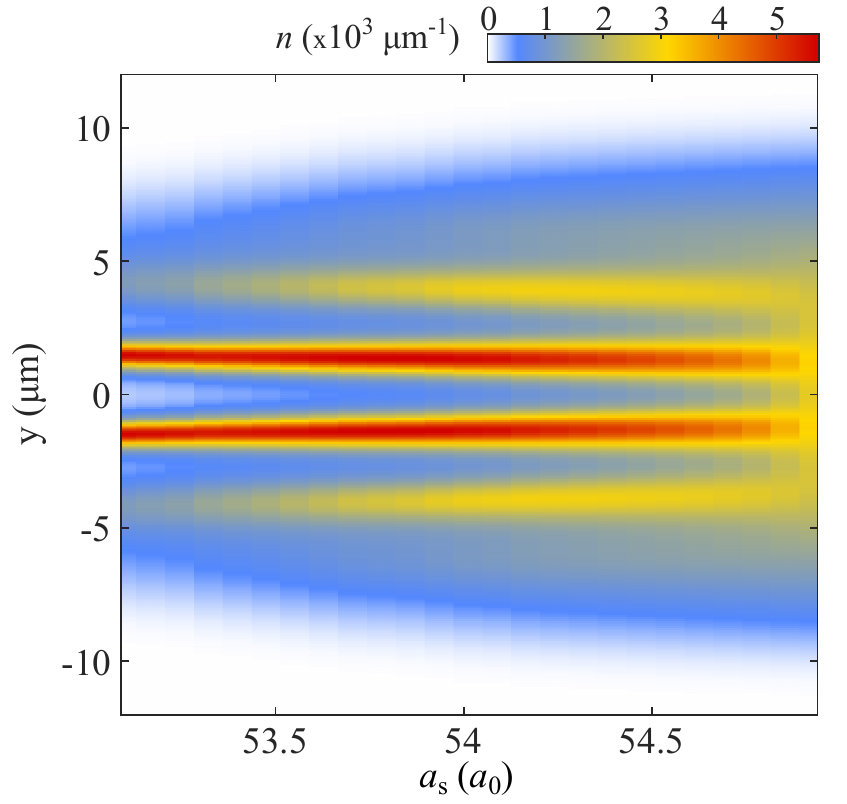}
	\caption {Radially integrated in-situ density profiles versus $\as$ of the supersolid ground states in a trap with harmonic frequencies $\omega_{x,y,z}=2\pi\times(250,31,160)$\,Hz. For each $\as$, we use the atom number measured in the experiment (see Fig.\,\ref{figS3:atomNumberVSHoldTime}, triangles).}
	\label{figS12:GroundstateDensVSas}
\end{figure}

For the current manuscript, we employ similar BdG and ground state calculations as already described in Refs.\,\cite{Chomaz:2018,Petter:2019,Natale:2019}. In this theory the gas is trapped in all three dimensions. For calculating the ground states, we use the experimentally extracted atom number at the intermediate timing of the Bragg pulse (presented in Fig.\,\ref{figS3:atomNumberVSHoldTime}, triangles). The radially integrated density profiles of the ground states in the SSP regime are presented in Fig.\,\ref{figS12:GroundstateDensVSas}. We note that in this theory, the BEC-SSP phase transition lies at $3.79\,a_0$ below the experimentally determined one. This shift in $\as$ between theory and experiment has also been found in Refs.\,\cite{Petter:2019,BoettcherQCorr:2019, Chomaz:2019}. Therefore, throughout the manuscript, the BdG theory and the ground state calculations are presented with an up-shift of $3.79\,a_0$ for $\as$. 

The presence of an axial trapping potential, leads to discrete excitation modes in the spectrum (typical energy spacing $\sim h\times20\,$Hz). Furthermore, each mode is broadened in $k_y$.
The finite duration of the Bragg pulse gives an energy broadening of each excitation mode (Fourier broadening $\sim h\times130\,$Hz full width at half maximum) which is much bigger than the energy spacing between the modes in the spectrum. Therefore, in the Bragg spectroscopy only a single resonance is visible, which is constituted of multiple excitation modes.
To account for this, we calculate $\DSF$ while broadening each mode in energy according to the Fourier-broadening, expected from a 7\,ms Bragg pulse. After calculating the broadened $\DSF$ and evaluating it at the experimental $\kBragg$, we also find in the BdG theory a single resonance in energy. To compare with the experiment, we extract $\DSFsp$ from a Gaussian fit to this resonance; see also~\cite{Petter:2019}.

\subsection{I. Free particle regime in the BdG theory for a trapped gas}
\begin{figure}[ht]
	\includegraphics[width=\columnwidth]{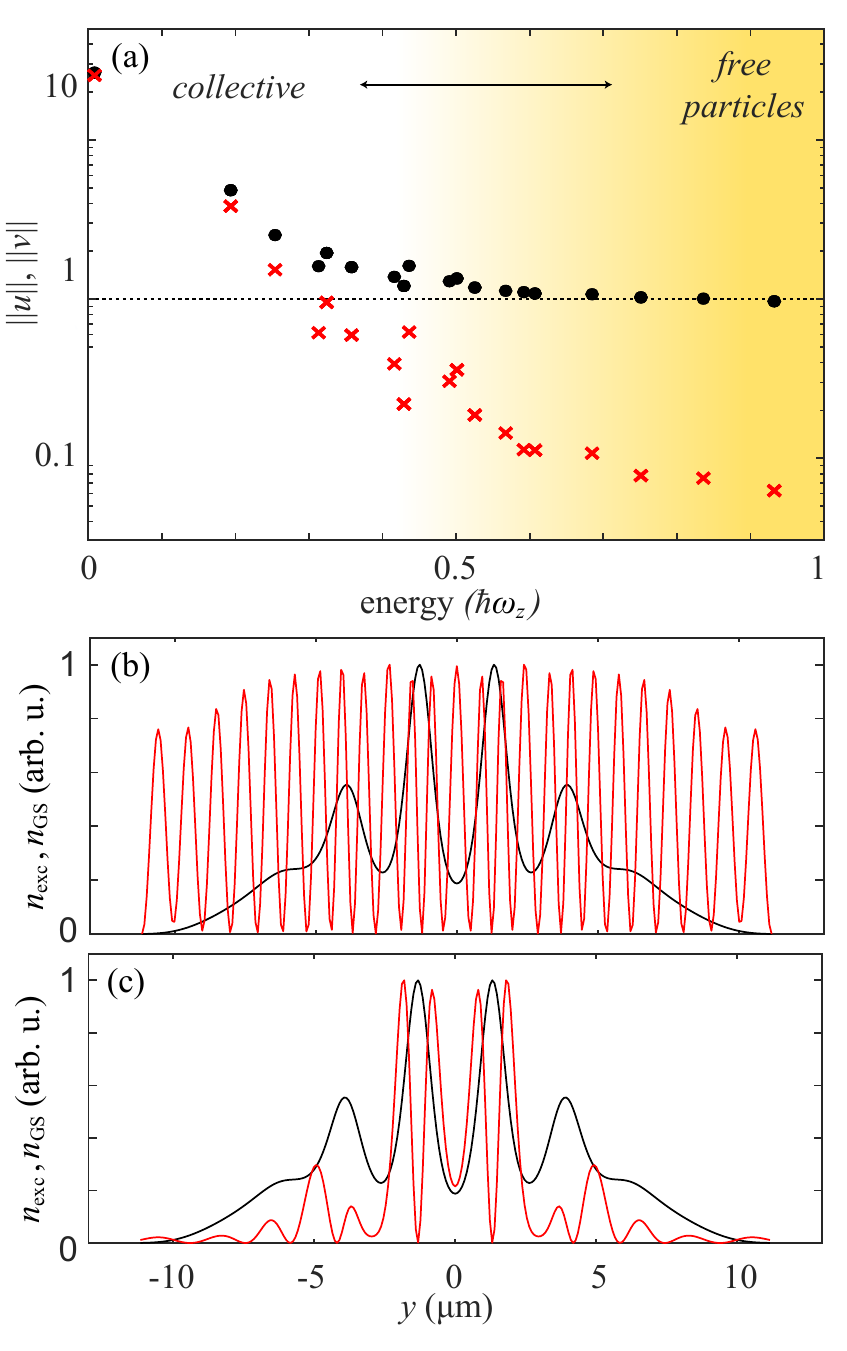}
	\caption {(a) Bogoluibov amplitudes, $\left\lVert v\right\rVert$ (crosses) and $\left\lVert u\right\rVert$ (circles), for the excitation modes of a trapped supersolid at $\as=54.49\,a_0$.
	(b) Integrated density profile, $n_\mathrm{exc}$, of an exemplary excitation mode with an energy of $1.7\,\hbar\omega_z$ (red line) and $k\approx4.1\,\upmu\mathrm{m}^{-1}$ and a comparison with the integrated ground state density, $n_\mathrm{GS}$, (grey line). (c) The same as in (b) but for an excitation mode at $0.55\,\hbar\omega_z$.
	}
	\label{figS4:FreeParticleRegime}
\end{figure}

To transfer the insights from the BdG calculations of an infinitely extended system (see Sec.\,G) to the experimentally trapped case, we also analyze $\left\lVert u\right\rVert$ and $\left\lVert v\right\rVert$ of the excitation modes obtained from the BdG calculations of a three-dimensional trapped gas.
Similar to the infinitely extended system, we find that modes with $\hbar\omega\gtrsim0.5\,\hbar\omega_z$ have $\left\lVert u\right\rVert\approx1$ and therefore a single-particle character across the BEC-SSP-ID regimes. This is exemplified in Fig.\,\ref{figS4:FreeParticleRegime}\,(a) where we show, for an exemplary state in the SSP, the norm of the obtained Bogoliubov amplitudes versus the energy of the corresponding mode.

As mentioned already in Sec.\,G, to further identify a single particle excitation as a free particle one, the excitation's wave function needs to be a plane wave. To investigate this aspect, we study the excitation modes' density profiles and find for modes in the experimentally relevant energy regime a clear plane wave character. Figure~\ref{figS4:FreeParticleRegime}\,(b) shows the radially integrated density profile of an exemplary excitation mode of a supersolid and compares it to the integrated density of the ground state. The plane wave character is clearly visible as a modulation with $k\approx4.1\,\upmu\mathrm{m}^{-1}$ across the whole system. We only find a mild reduction of the plane wave's amplitude towards the outer region of the system.
As a comparison, we show in Figure~\ref{figS4:FreeParticleRegime}\,(c) the integrated density profile of an excitation mode at lower energy, $0.55\,\hbar\omega_z$, which also has $\left\lVert u\right\rVert\approx1$, but is clearly not a plane wave.

\subsection{J. Comparison of the finite-size BdG theory to the self-consistent SIA and DAA model}
From the ground state profiles of the three-dimensional trapped BdG theory in Fig.\,\ref{figS12:GroundstateDensVSas}, we numerically evaluate our SIA and DAA model on the corresponding ground states. 
To self-consistently evaluate the SIA result from the ground state, we need to estimate the contrast of the density modulation, $C=(n_\mathrm{max}-n_\mathrm{min})/(n_\mathrm{max}+n_\mathrm{min})$. We determine $n_\mathrm{min}$ from the minimum density at $y=0$ and $n_\mathrm{max}$ from the density of one of the two most-central density peaks. To evaluate the DAA model, we numerically extract the $1/\mathrm{e}$-size, $\sigma$, of the two central droplets and the distance $d$ between them. To estimate the density, we  calculate the mean density in the central region between the two central droplets. Therefore, our model comparison takes only the central part of the system into account and neglects the outer density regions.
Furthermore, as our models extract only the scalings of the DSF with the ground state properties and not its absolute value, we renormalize the SIA and the DAA. For the comparison in Fig.\,1 of the main manuscript we show the finite BdG and the SIA rescaled to unity for the point at the phase transition. The DAA is rescaled directly on the BdG theory to match its values in the lower $\as$ regime. Over the investigated $\as$ range, we find that both, the SIA and the DAA, describe the BdG theory well for low and large $C$, respectively (see main text), and  for momenta $k\geq4.0\,\upmu \mathrm{m}^{-1}$. This gives an estimate for which momenta the impulse approximation becomes valid.

\subsection{K. Real time evolution of the Bragg scattering}
Our theory for the RTE simulations was already presented in Ref.\,\cite{Chomaz:2018}. For the current manuscript, the simulations start with the ground state wave function with $N=8.5\times10^4$ atoms at $\as=60.9\,a_0$. We add a thermal population, corresponding to a randomly drawn occupation of the system's excited states (incl. a random phase) with a Poisson distribution, whose mean is given by the Bose distribution ($+1/2$ to simulate quantum fluctuations) for the mode's energy at a temperature of 100\,nK~\cite{Blakie:2008}. This increases the total atom number to about $1\times10^5$ (similar to the experimental situation) and simulates thermal and quantum fluctuations in the system.

In the time evolution, we reproduce the experimental sequence, including a 20\,ms long linear $\as$ ramp, followed by a 10\,ms holding at the final $\as$ and a consecutive 7\,ms Bragg excitation along $y$. In order to obtain the system's momentum distribution, we perform a Fourier-transform of its wave function. On this momentum distribution, we perform the same analysis as on the experimental data, i.\,e. we analyse the fraction of excited atoms in a region of interest around $\kBragg\approx4.2\,\upmu\mathrm{m}^{-1}$ for various $\omega$. The resonances are fitted with a Gaussian function to obtain $\fexc$ from the RTE simulations.
For the same reasons as mentioned for the BdG calculations (Sec.\,H.), the crossing of the BEC-SSP phase transition in the RTE happens $3.34\,a_0$ below the experimental phase transition. Therefore, throughout the manuscript, the RTE theory is up-shifted in $\as$ by $3.34\,a_0$.
We note that, when performing the RTE directly on the corresponding ground states from the BdG theory, i.\,e. we do not include thermal noise, the $\as$-ramp and atom loss, we recover an excited fraction that is well described by the BdG theory. This indicates that the chosen analysis in ToF gives a reliable measurement of $\DSFsp$ and in particular a consistent result with the 3D-trapped BdG calculations.

\subsection{L. Real-time evolution and characteristics of the state's wave function}

To study the time evolution of the contrast and the phase of the dynamically created supersolid states, we perform RTE simulations without applying a Bragg excitation and monitor the axial density and the phase profiles from the calculated wave functions [see Fig.~4\,(a,\,c) in main manuscript]. Typically, in the RTE we observe $N_D=4-6$ droplets, containing a variable atom number across the system. We evaluate the time-dependent central contrast, $C$, between the two central density peaks numerically. The phase-variation factor $\Delta\Theta=|\frac{1}{N_\mathrm{D}}\sum_{j=1}^{N_D}\langle e^{i\theta_j}\rangle_\tauBragg|^2$ is calculated by extracting the mean phase, $\theta_j$, of each single density peak over its full-width at half maximum.

\begin{figure}[ht]
	\includegraphics[width=\columnwidth]{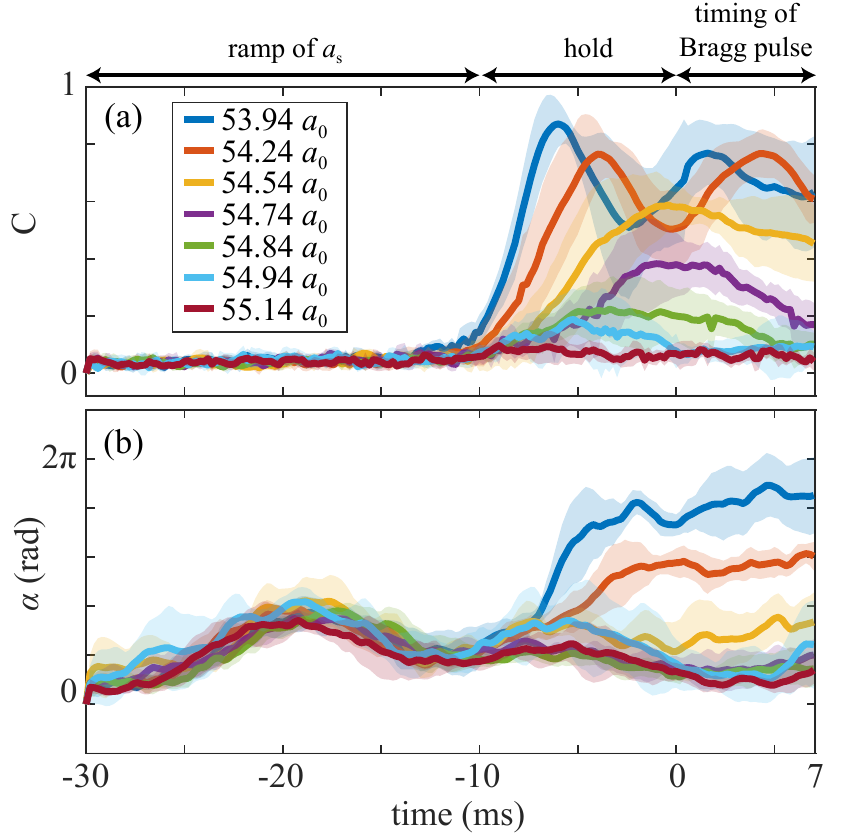}
	\caption {RTE simulations with $V=0$ (no Bragg excitation applied). (a) The time evolution of the extracted central contrast of the integrated in-situ density distributions for different $\as$ (see legend). (b) The extracted phase incoherence of a central cut through the wave function (see text). The shadings represent the standard deviation from 5 simulation runs with different statistical draws of the thermal population. 
    The data presented in Fig.\,\ref{fig4:timeEvolRTE}\,(a,\,b) corresponds to the time window of $[0,\,7]\,$ms.}
	 \label{FigS9:PhaseIncoherence}
\end{figure}

Figure~\ref{FigS9:PhaseIncoherence}\,(a) shows $C(t)$ over the whole simulation time. For early times, we observe a small, but finite $C$ due to density noise in the simulations, which is coming from the included thermal fluctuations. During the holding time ($[-10,\,7]\,$ms), for $\as<\ascritExp$, we observe that $C$ first increases and consecutively slightly decreases due to atom loss. Only for $\as<54.2\,a_0$ we find an oscillating behaviour of the contrast in time. We note that there is a time delay between the development of the density modulation in the system and the timing of the $\as$ ramp (which occurs during $[-30,\,-10]\,$ms).

To give another insight into the time evolution of the system's phase profile, we extract the global phase variation, $\alpha=\frac{1}{L}\int_{L}|\phi(0,y,0)-\braket\phi_{L}|dy$, of the wave function, which is extracted along a cut of $\phi$ along $y$. Here, $\braket\phi_{L}$ denotes the averaged phase over the central region $L =[-7,7]\,\upmu$m, see also \cite{Tanzi:2019}. In Figure~\ref{FigS9:PhaseIncoherence}\,(b), we show the time evolution of $\alpha$ for the whole simulation time. For all $\as$ one sees a first local maximum in $\alpha$ (around $-20\,$ms), coming from an axial breathing mode which is excited due to the $\as$ ramp. At longer times, we find for $54.5\,a_0\lesssim\as \leq\ascritExp $, that $\alpha$ remains small while the density contrast is finite. For $\as<54.5\,a_0$, we find that $\alpha$ seems to approach a constant value, which increases with lower $\as$.

\end{document}